\documentstyle[twoside,fleqn,espcrc2,epsf]{article}

\newcommand{\AmS}{{\protect\the\textfont2
  A\kern-.1667em\lower.5ex\hbox{M}\kern-.125emS}}

\newcommand{\noi}{\noindent}
\newcommand{\eq}{\begin{equation}}
\newcommand{\en}{\end{equation}}
\newcommand{\eqa}{\begin{eqnarray}}
\newcommand{\ena}{\end{eqnarray}}

\newcommand{\inta}{\int_{A}~}

\newcommand{\aleq}{\mbox{}_{\textstyle \sim}^{\textstyle < }}
\newcommand{\ageq}{\mbox{}_{\textstyle \sim}^{\textstyle > }}

\title{
\vspace{-1.8cm}
\hbox{}
{\small NOVEMBER 1994} \hfill {\small HU BERLIN--IEP--94/28~}   \break
\hbox{}                \hfill {\small IFUP--TH~~71/94}          \break
                                                              \break
On $T$ dependence of the static potential $V(T;{\vec R})$
in a finite volume
\thanks{TALK GIVEN AT THE LATTICE '94 INTERNATIONAL SYMPOSIUM
       LATTICE FIELD THEORY, BIELEFELD, GERMANY,
       SEPTEMBER 27 -- OCTOBER 1, 1994}
\thanks{Work supported by the Deutsche
Forschungsgemeinschaft under research grant Mu 932/1-3 }
}

\author{G.~Cella\address{
INFN in Pisa and Dipartimento di Fisica dell'Universit\'a
di Pisa,  Italy },
V.K.~Mitrjushkin\address{ Institut f\"{u}r Physik, Humboldt-Universit\"{a}t,
10099 Berlin, Germany}
\thanks{Permanent adress:
JINR, Dubna, Russia}
and
A.~Vicer\'e$\mbox{}^{\scriptsize a}$
}

\begin{document}

\begin{abstract}
We study the dependence on $T$ of the static potential $V(T;{\vec R})$
defined from Wilson loops and from Polyakov loop correlators on a finite
lattice.  For this study we employ a simple model with
confinement, and compare with MC results.
\end{abstract}

\maketitle

\section{Introduction}

The usual way to define the potential on a finite lattice
with $N=L_t L_s^3$ is

\eq
V_{W}(T;{\vec R}) = \ln \frac{W(T;{\vec R})}{W(T+1;{\vec R})}
\en

\noi where $~W(T;{\vec R})~$ is the average of the Wilson loop with 'time'
extension $~T~$ and space extension $~R~$. The Wilson loop can be
on--axis or off-axis, and the space--like parts of the loop
can include the contribution of many different contours.
It is common to use the smearing procedure \cite{ape} to improve the
signal--to--noise ratio.

In the confinement phase the potential $V_W(T;{\vec R})$ obtained in
such a way
shows typically the following features :\\
\noi a) it linearly rises up to the distances $~R \ageq L_s/2~$;\\
\noi b) it is (approximately) rotationally invariant {\it even}
at the distances $R \sim L_s/2$;\\
\noi c) it shows rather weak dependence on $T$ at $T \sim 4 \div 6$
(when smearing was used). This usually gives the reason to believe that
$T=\infty$ limit is under control.

Another way to define the potential is to use the Polyakov loop
correlator $~\Gamma_{P}({\vec R})~$

\eq
V_{P}({\vec R}) = -\frac{1}{L_t} \cdot
 \ln \Gamma_{P}({\vec R}).
                                      \label{v_p1}
\en

In numerical simulations on finite lattices both definitions give
usually different values for the string tension, and some efforts are
necessary to make them consistent (see, e.g., \cite{gao}).

It is the aim of our study to make a {\it model} analysis of
the static potential $V$ on a finite lattice
for both definitions with a special accent on the large $T$ behaviour
of $V_W$.

\section{Coulomb phase}

It could be instructive first to make few comments about the
perturbative potentials.  At small enough couplings $g^2$ the standard
perturbation expansion can be applied to calculate both $V_{W}$ and
$V_{P}$. For Polyakov loop potential $V_P$ one obtains the 'lattice Coulomb'

\eq
V_{P}({\vec R}) = \frac{4\pi}{L_s^{3}}
\sum_{{\vec p}} ~ (1 - e^{i {\vec p} {\vec R}})
 G_{coul} (p_4=0;{\vec p})~;
                                      \label{v_p2}
\en

\noi which becomes $\sim 1/R$ (up to the constant) on the
{\it infinite} lattice.
Here $G_{coul}^{-1}(p) =\sum_{\mu =1}^{4} 4\sin^2 \frac{p_{\mu}}{2}$.

The Wilson loop potential $V_{W}$ can be represented (in Feynman
gauge) as $ V_{W} \equiv V^{t}_{W}+ V^{s}_{W} $ where
\eq
V^{t}_{W} \equiv Q^t(T;{\vec R}) - Q^t(T+1;{\vec R})
                      \label{v_t1}
\en
\noi is the time--time ('electric') contribution with

\eq
Q^{t} = \frac{4\pi}{N } \sum_{p} ( 1 - e^{i {\vec p}{\vec R} } )
\frac{1 - \cos p_4T}{1 - \cos p_4} \cdot G_{coul}(p)~,
                      \label{v_t2}
\en

\noi and
\eq
V^{s}_{W} \equiv Q^s(T;{\vec R}) - Q^s(T+1;{\vec R})
                      \label{v_s1}
\en
\noi is the space--space ('magnetic') contribution with
\eq
Q^{s} = \frac{4\pi}{N } \sum_{p} (1 - \cos p_4T)
\frac{1 - \cos p_3R }{1 - \cos p_3 } \cdot G_{coul}(p).
                      \label{v_s2}
\en

\noi In eq.(\ref{v_s2}) the planar Wilson loop in the plane
$~(x_4;x_3)~$ was chosen. The generalization to the off--axis Wilson
loops is straightforward. In the limit $T \rightarrow \infty$
(on a lattice $\infty \cdot L_s^3$) the
'magnetic' part $V_W^s$ tends to zero, and $V_W^t \rightarrow V_P$.

Some useful observations could be made here.\\
\noi a) 'Lattice Coulomb' $V_P$ differs {\it significantly}
from its continuum counterpart for lattice sizes typical for MC
calculations (i.e., $L_s \sim 16 \div 32$). In fact, the usage of
the continuum $\sim 1/R$ potential in fit formula on {\it finite}
lattices is not justified.\\
\noi b) At small $T$ the 'magnetic' part $V_W^s$ is {\it not} small and
gives dominant
contribution to the Wilson loop potential $V_W$ at $R \sim L_s/2$
(which grows with $R$).
It becomes small enough only  when $T \sim L_s$ (i.e., elongated lattice
with $L_t \ageq 2L_s$ is needed).\\
\noi c) The 'electric' part $V_W^t$ in its turn deviates from $V_P$
for small $T$ and $R \sim L_s/2$.  Again, they become close enough to each
other only when $T \sim L_s$.

This exercise shows that some care is necessary in
analysis on a finite lattice (see also \cite{em}).

\section{Confinement phase}

The {\it model} we have chosen here is based on the representation of the
partition function $Z$ in the following form

\eq
Z = \inta e^{-S{ef\! f}(A) }
\equiv \inta e^{- g^2 \kappa^{-1} \cdot A G_{conf}^{-1} A}~,
\en

\noi with the confining--type propagator
$G_{con\! f}^{\mu \nu}(p) = \delta_{\mu \nu} \cdot G_{con\! f}(p)$

\eq
G^{-1}_{con\! f}(p) =
16 \cdot \Big[ \sum_{\mu =1}^{4} \sin^2 \frac{p_{\mu}}{2} \Big]^2,
\en

\noi where $\kappa =\kappa (g)$ is some constant.
The zero--momentum mode ($p = 0$) is excluded.

The Polyakov loop potential $V_P$ is

\eq
V_{P}({\vec R}) = \frac{8\pi}{L_s^{3}}
\sum_{{\vec p}} ~ (1 - e^{i {\vec p} {\vec R}}) G_{conf} (p_4=0;{\vec p})~;
                                      \label{v_p3}
\en

\noi The continuum analog of $V_P$ is the linearly rising potential

\eq
V_{cont}({\vec R}) = \frac{1}{\pi^2} \cdot
\int d {\vec p} ~ \frac{1 - e^{i {\vec p} {\vec R}}}{ ({\vec p}^{\, 2} )^2}
\equiv R ~.
                                      \label{v_con1}
\en

\noi On finite lattices $V_P$  deviates strongly from
 $V_{cont} =R$. The infinite volume
behaviour can be reproduced only at $1 \ll R \ll L_s/2$, which needs
lattice sizes $L_s \sim 10^2 \div 10^3$.

For Wilson loop potential $V_W=V_W^t+V_W^s$ one obtains
the expressions similar (up to normalisation) to that in
eq.(\ref{v_t1}) $\div$ eq.(\ref{v_s2}) with the only difference that
instead of $G_{coul}(p)$ enters $G_{con\! f}(p)$.

 The dependence of $V_W(T;{\vec R})$ on $R$ at different $T$
on $32 \cdot 16^3$ lattice is
shown in Fig.1.  The interesting observation is that for $R \sim L_s/2$
very {\it weak} dependence on $T$ takes place for $T \aleq 8\div 10$,
i.e., there is a 'plateau' in the $T$-dependence.  (It is worthwhile
to recall here that in MC calculations the smaller values of $T$ are
usually employed).  With increasing $T$ ($T >10$) this plateau disappears,
and $V_W$ differs significantly from the potential at small $T$'s.

The details of this picture depend on the ratio of $L_t/L_s$, and
the role of the static modes ($p_4 \neq 0;~{\vec p} = 0$) still needs
some clarification.

In Fig.2 we show the $R$ dependence of $V_W$ and its 'electric' and
'magnetic' components at $T=4$. The off--axis Wilson loops are included.
The 'electric' part which is supposed to give at big $T$ the correct
potential of two static charges is far from being linear.
The 'magnetic' part dominates at small $R$, and even at $R \sim L_s/2$
its contribution is rather big.
The potential itself shows the linearly rising behaviour and
the (approximate) rotational invariance
(which disappear at larger $T$'s).

\begin{figure}[htb]
\begin{center}
\vskip -1.8truecm
\noi
\leavevmode
\hbox{
\epsfysize=300pt\epsfbox{fig1.ps}
     }
\end{center}
\vskip -3.5truecm
\caption{
The dependence of potential $V_W(T;{\vec R})$ on $R$
on a $32 \cdot 16^3$ lattice for different values of $T$.
Broken line corresponds to $V_P({\vec R})$.
}
\label{fig:v_w1}
\vskip -0.9truecm
\end{figure}

\begin{figure}[htb]
\begin{center}
\vskip -1.8truecm
\noi
\leavevmode
\hbox{
\epsfysize=300pt\epsfbox{fig2.ps}
     }
\end{center}
\vskip -3.5truecm
\caption{
$V_W(T;{\vec R})$, $V_W^t(T;{\vec R})$ and  $V_W^s(T;{\vec R})$
as a function of $R$ at $T=4$ on a $~32 \cdot 16^3~$ lattice.
}
\label{fig:v_w2}
\vskip -0.7truecm
\end{figure}

\begin{figure}[htb]
\begin{center}
\vskip -1.7truecm
\noi
\leavevmode
\hbox{
\epsfysize=300pt\epsfbox{fig3.ps}
     }
\end{center}
\vskip -3.5truecm
\caption{
The comparison of the  $U(1)$ MC potential
$V_{W}(T;{\vec R})$ without smearing with $V_{fit}(T;{\vec R})$
on a $~32 \cdot 16^3~$ lattice.
}
\label{fig:fit}
\vskip -0.7truecm
\end{figure}


Our model permits to obtain a reasonable fit of MC data.
In Fig.3 we show the results of the fit of $V_W(T;{\vec R})$
obtained from numerical calculations  without smearing.
The fitting potential is
\eq
V_{fit}(T;{\vec R}) = C + \alpha \cdot V_{coul}
 + \sigma \cdot V_{conf}
                   \label{vfit}
\en

\noi where parameters of the fit $C$, $\alpha$ and $\sigma$ do not
depend on $T$. We used here the $T$--dependent 'lattice Coulomb'
potential $V_{coul}(T;{\vec R})$ from eq.(\ref{v_t1}) $\div$
eq.(\ref{v_s2}), and similar expressions for $V_{conf}(T;{\vec R})$.
At least qualitatively the model reproduces the main features
of the Wilson loop potential.

\vskip -5mm
\section{Conclusions}

Our confining--type model reproduces the main features of the
static potential defined from Wilson loops at small $T$.  The
comparison with MC data without smearing shows that
a reasonable (at least, qualitatively) fit can be obtained.

The main conclusion is that it {\it may} appear to be dangerous to
extract the static potential at comparatively small $T$.
One can not exclude that with increasing $T$ the potential
$V_W(T;{\vec R})$ will become drastically different.

At small $R$ the {\it finite} $T$ 'lattice Coulomb'
should be used for the fit of the {\it finite} $T$ Wilson loop
potential (at least, for the unsmeared data).

We believe that this topic deserves additional
study (both numerical and analytical).
Also, the role of the smearing should be studied more.

\end{document}